\documentstyle[12pt]{article}
\newcommand {\iMi}{\int d^4x \sqrt {-g}}
\newcommand {\cD}{{\cal D}}
\newcommand {\gM}{\sqrt {-g}}
\newcommand {\oB}{\vert_{\partial M}=0}
\newcommand {\abL}{\vert l \vert}
\begin{document}

IC/94/359

gr-qc/9411036

\centerline{\large \bf QED on Curved Background and on Manifolds
with Boundaries:}
\centerline{\large \bf Unitarity versus Covariance}
\vspace {15mm}
\centerline{Dmitri V.Vassilevich \footnote {Permanent address:
Department of Theoretical Physics, St.Petersburg University,
198904 St.Petersburg, Russia, e.-mail: dvassil @ sph.spb.su}}

International Centre for Theoretical Physics, Trieste, Italy

\vspace {20mm}

\centerline{Abstract}

Some recent results show that the covariant path integral and
the integral over physical degrees of freedom give contradicting
results on curved background and on manifolds with boundaries.
This looks like a conflict between unitarity and covariance.
We argue that this effect is due to the use of non-covariant
measure on the space of physical degrees of freedom. Starting
with the reduced phase space path integral and using covariant
measure throughout computations we recover standard path
integral in the Lorentz gauge and the Moss and Poletti
BRST-invariant boundary conditions. We also demonstrate by
direct calculations that in the approach based on Gaussian
path integral on the space of physical degrees of freedom
some basic symmetries are broken.


\newpage
{\bf 1. Introduction}

A self-consistent approach to quantization of gauge theories was
suggested more than a quarter of century ago in a pioneering paper
by Faddeev and Popov [1]. In this and subsequent works [2] it
was demonstrated that covariant gauge quantization rules can be
derived from manifestly unitary reduced phase space approach.
A suitable quantization procedure for the case of curved space-time
was also elaborated (see [3,4,5]). However contradictions between
covariant approach and Hamiltonian approach were reported for
gravity and electrodynamics on de Sitter space [6-11] and on manifolds
with boundaries [5.12-18]. This is a very serious problem since there is no
use of a theory which is non-covariant or not unitary. Proper
field theory should admit some kind of probabilistic interpretation
and have all fundamental symmetries conserved, i.e. it should
posses both unitarity and covariance properties.

Some progress was achieved in overcoming the above mentioned
difficulty for unbounded manifolds . It was demonstrated [19]
that the path integral over physical degrees of freedom of
quantum gravity is equivalent to the covariant result by
Taylor and Veneziano [7] provided covariant functional measure
is used throughout computations. A correct path integral measure
for Hamiltonian QED on Friedmann-Robertson-Walker background was
constructed [20]. It was suggested [21] that the discrepancy
between the two approaches is due to the fact that on certain
manifolds the 3+1 decomposition is ill-defined. It was demonstrated
[22] that on a flat space region between two concentric spheres
the value of
the scaling behavior $\zeta (0)$ is gauge independent.
However, to obtain this result it was necessary to include
quantization over non-physical degrees of freedom and ghosts.
Some recent works show
[23,24] that a part of difficulties of quantum field theory on
manifolds with boundaries, but not all of them, are due to
mistakes in calculations.

Before going into details, let us give some motivations.
Consider two path integrals, one over covariant configuration
space and the other over physical degrees of freedom (PDF).
Suppose, that one of these two path integrals satisfy both
covariance and unitarity requirements. Let us try to guess which
one most probably gives correct results. It was not demonstrated
anywhere that the covariant path integral violates unitarity.
Let us check up covariance of the PDF path integral. To this end
let us recall the heat kernel expansion for electrodynamics
on unit four-sphere [11] and unit radius four-disk [13].
$$W_{\rm PDF}=-\log \int_{\rm PDF} \cD A \exp (-S(A))=
-\frac 12 \int_0^\infty \frac {dt}t K(t)$$
$$K(t)[S^4]=\frac 1{3t^2}-\frac 1t +\frac 12
\sqrt {\frac {\pi}t }-\frac {16}{45} +O(\sqrt t )
\eqno (1.1)$$
$$K(t)[{\rm Disk}]=\frac 1{16t^2} -\frac {\sqrt \pi}{8t^\frac 32}
+\frac 1{\sqrt t} \left ( \frac {53 \sqrt \pi}{256} +
\frac 1{2\sqrt \pi} \right ) -\frac {77}{180} +O(\sqrt t )
\eqno (1.2)$$
An experienced reader would see what is wrong with equation (1.1).
The heat kernel expansion of a covariant theory on a closed manifold
can not contain term with half integer powers of proper time.
Such a term leads to divergency which can not be cancelled by
local covariant counter-term. This indicates that the covariance
property is lost in the PDF formalism on $S^4$. The situation
with the expression (1.2) is more subtle. On manifolds with
boundaries terms with half integer powers of $t$ are allowed.
However, looking at the general form of the heat kernel
expansion [25], we see that on the unit disk the $\sqrt \pi$
naturally appears in numerator, but not in denominator as in
one of the terms of (1.2). Such terms with $\pi^{-\frac 12}$
are not strictly forbidden, but their appearance is quite
unsatisfactory. These terms hardly can be reproduced in
a covariant theory. We see that in the PDF formulation
covariance is most probably violated. Thus the covariant approach
is better candidate for a proper theory than the PDF one.

The present paper is devoted to the study of QED on curved
space and on manifolds with boundaries paying special attention
to covariance and unitarity issues. First we shall derive
covariant path integral from the reduced phase space quantization
(sec. 2) using covariant path integral measure. This derivation
shows that the source of differences between two approaches
was in using non-covariant measure on reduced phase space.
The proper measure on the space of physical degrees of freedom
appeared to be complicated. This fact should not be a surprise
(see [26]). We extend the phase space in order to be able to
use covariant measure  for four-vector fields. We also
consider an important particular case of static space-time.
In this section we assume that the problems with boundary
conditions are somehow regulated. In the next section 3 we
give several examples. We demonstrate the loss of invariance
properties in PDF formalism for electrodynamics on torus and
sphere. We also re-calculate some results connected with
criterion of normalizability of the wave function of the
Universe using covariant technique. The section 4 is devoted
to manifolds with boundaries. We demonstrate that the procedure
of section 2 leads to the Moss and Poletti boundary conditions
[14,27] for electrodynamics. We also re-derive these boundary
conditions using manifestly covariant procedure. Appendix A
contains some additional illustrative material to sec. 3.
Useful relations of differential geometry of manifolds with
boundaries are collected in the Appendix B.

Now some definitions are in order. A path integral measure
$\cD v$ on the space $V$ with scalar product $<\ ,\ >$
will be called Gaussian if
$$\int_V \cD v \exp (-<v,v>)=1 \eqno (1.3)$$
The measure $\cD v$ will be called covariant if both the
integration space $V$ and scalar product $<\ ,\ >$ posses
all necessary invariance properties. In our case these are
the Lorentz and diffeomorphism invariance.

The definition (1.3) is written for Euclidean path integral.
In the Minkowski signature space the minus sign in
exponent should be replaced by $i$. These definitions
follow the approach by DeWitt [3] and Polaykov [28].

\newpage
{\bf 2. Quantization and the measure}

Consider the action for electromagnetic field $A_\mu$
$$S=-\iMi \frac 14 F_{\mu \nu}F^{\mu \nu},
\quad F_{\mu \nu}=\partial_\mu A_\nu -\partial_\nu A_\mu
\eqno (2.1)$$
On a Friedmann-Robertson-Walker background ($g_{0i}=0$,
$g_{00}=-N^2=$const) one can rewrite (2.1) in the form
$$S=\int dt \int d^3x [\sqrt {(-g)} P^i (\partial_0 A_i)-
({\cal H} +\lambda \Phi )] \eqno (2.2)$$
where $\sqrt {-g}P^i=\sqrt {-g}F^{i0}$ is the momentum
conjugate to $A_i$, and
$$H=\int d^3x {\cal H}=\int d^3x \sqrt {-g} \frac 14
[2 P_iP^i N^2 +F_{ik}F^{ik}] \eqno (2.3)$$
plays the role of Hamiltonian. Last term in (2.2) generates
the first class constraint
$$\Phi =-^{(3)}\nabla_i P^i=0, \eqno (2.4)$$
where $\ ^{(3)}\nabla$ is the covariant derivative with
the respect to three-metric $g_{ik}$. To obtain (2.3)
an integration by parts over $x_j$ is needed. We assume
that either spatial sections are closed manifolds or
all the fields decay rapidly at spatial infinity.
The problem of the boundary conditions in the time
direction is more subtle. It will be considered
in the section 4.
The canonical
Poisson bracket is
$$\{ A_i (x,t), \sqrt {-g}P^k (y,t) \}=\delta^3 (x-y)
\delta^k_i. \eqno (2.5)$$
The constraint (2.4) generates gradient transformation
of $A_i$. The corresponding gauge freedom can be fixed by
imposing the condition
$$\chi=\ ^{(3)}\nabla_i A^i=0 \eqno (2.6)$$
Note that in general the condition (2.6) does not fix the
gauge freedom of the four-dimensional action (2.1) completely.
The transformations
$$\delta A_\mu =\partial_\mu \omega (t) \quad
{\rm with}\quad \partial_i \omega (t) =0 \eqno (2.7)$$
do not change canonical coordinates and thus are allowed by
(2.6). If a spatial section is homeomorphic to $R^3$, the
transformations (2.7) are excluded by conditions at spatial
infinity. However, these transformations are allowed for
compact spatial sections. We shall return to this problem
a bit latter.

As usual, the reduced phase space can be obtained by
solving equations (2.4) and (2.6). This space consists
of transverse spatial components of vector potential,
$A_i^T$, and transverse components of conjugate
momentum, $P^{Tj}$. The canonical bracket (2.5) takes the form
$$\{ A_i^T (x,t), \sqrt {-g} P^{Tk} \} =
\left ( \delta_i^k -
\frac {\ ^{(3)}{\nabla_i}^{(3)}\nabla^k}{\ ^{(3)}\Delta}
\right ) \delta (x-y) \eqno (2.8)$$

We see that the right hand side of (2.8) is now non-
trivial. This means that the measure $\cD A_i^T \cD P^{Tk}$
should include the determinant of the functional metric
induced on the surface defined by the conditions (2.4)
and (2.6). This determinant depends upon geometry of
space-time manifold, boundary conditions, etc. The evaluation
of this determinant is highly complicated task.
It is much easier to extend the phase space in order to
be able to work with standard inner product of four-vectors
$$<u,v>=\iMi v_\mu u^\mu \eqno (2.9)$$
According to the general prescription [1,2] the path integral
can be written in the form
$$Z=\int \cD A_i \cD \sqrt {-g} P^k \cD \lambda
\delta (\chi ) \det \{ \chi ,\Phi \} \exp (-iS) \eqno (2.10)$$
with the action (2.2). The Lagrange multiplier $\lambda$ is
just the weighted zeroth component of the vector potential,
$\lambda=$$\sqrt {-g} A_0$. The fields $A_0$ which do not
depend on spatial coordinates do not generate any constraints.
Thus one should exclude spatial constants from the integration
measure $\cD \lambda$. This can be done by means of the condition
$$\tilde \chi =\partial_0 \int d^3x \sqrt {-g} A_0 =
\partial_0 <\sqrt {-g} A_0>=0 \eqno (2.11)$$
Other choices are also possible.
{}From the point of view of the covariant theory, eq. (2.11) is nothing
else than fixation of the gauge freedom (2.7). This leads to
the appearence  of the Jacobian factor $\tilde J$ in
the integration measure
$$\cD \lambda =\tilde J \delta (\tilde \chi ) \cD A_0 \eqno (2.12)$$
The condition (2.11) appears naturaly [29] in a framework of
the BRST quantization. Previously it was demonstrated [20,29]
that $\tilde J=$$\prod_{x_i} \det (-\nabla_0^2)$.

Now we are to extend the phase space in order to obtain an
integral over four-vector corresponding to $P^k$. This can
be done by inserting the identity
$$1=\int \cD a \cD \gM P^0 \exp (-i\iMi [aP^0 -\frac 12
P^{02}N^4]
\eqno (2.13)$$
where $a$ is a scalar field.
After some algebra the path integral (2.10) is
represented in the form
$$Z=\int \cD A_\mu \cD \gM P^\mu \cD a \tilde J
\det \{ \chi , \Phi \} \delta (\chi ) \delta (\tilde \chi )
\exp (-i S(A_\mu)+$$
$$+\frac i2 \iMi \left ( (P_\mu -P_\mu (A)) (P^\mu -P^\mu (A))N^2
+ a^2 N^{-4}  \right ) \eqno (2.14)$$
where $S(A)$ is the action (2.1), $P^i(A)$ is the canonical
expression for $P^i$ in terms of $A_i$, $P^0(A)=a$.

The main advantage of the expression (2.14) is that now the
momentum integration is performed over unconstrained four-vector
$P^\mu$. We can use standard definitions of the covariant
Gaussian measures for vector density and scalar field
$$\int \cD \gM v^\mu \exp (i<v,v>)=1 \eqno (2.15)$$
$$\int \cD a \exp (i \iMi a^2 )=1 \eqno (2.16)$$
The integration over $a$ and $P^\mu$ produces ill-defined
factor $\prod_x N^{-2}$. This factor can be neglected
in dimensional or zeta-regularization because it does not contain
derivatives and does not depend on coordinates. It does not
contribute to $\zeta (0))$. Moreover, for Friedmann-Robertson-
Walker space-time $N$ can be absorbed in definition of time
variable. Important thing that $N$ enters path integral
in even power and hence does not lead to imaginary factor
after continuation to Euclidean space.
Using the definitions (2.15) and (2.16)
 we obtain the following expression for
generating functional $Z$ in terms of the functional integral
over $A_\mu$
$$Z=\int \cD A_\mu \delta (\chi ) \delta (\tilde \chi )
\det \{ \chi , \Phi \} \tilde J \exp (-iS(A_\mu )) \eqno (2.17)$$
This is the familiar form of the path integral of electrodynamics
with delta-functions of gauge conditions. One can remove
delta-functions by introducing gauge-fixing terms. However,
in this case, apart from ill-defined term $\tilde \chi^2$,
we shall also get non-covariant term $\chi^2$. This would lead
to non-covariant 3+1 splitting of $A_\mu$ and working with
non-covariant measures.

The most elegant way to evaluate the integral (2.17) using
only covariant measures is to introduce the so-called adapted
coordinates [30]. Let us make the Hodge-de Rham decomposition
of the four-vector $A_\mu$
$$A_\mu =A_\mu^\perp +A_\mu^\parallel ,\quad
A_\mu^\parallel =\partial_\mu \omega , \quad
\nabla^\nu A_\nu^\perp =0. \eqno (2.18)$$
This decomposition is orthogonal with respect to the scalar
product (2.9). The Jacobian factor due to the change
of variables $A_\mu \to$$(A^\perp ,\omega )$ is just
the determinant of the scalar four-Laplacian,
$\det (-\Delta )^{\frac 12}_S$ . The measure induced on
the subspace $A^\perp$ is Gaussian. Indeed,
$$\int \cD A_\mu^\perp \exp (i<A^\perp ,A^\perp >)=
\int \cD A_\mu^\perp \exp (i<A^\perp ,A^\perp >)\times $$
$$\times \int \cD \omega \exp (i<\omega ,(-\Delta )\omega >)
\det (-\Delta)^{\frac 12}_S= \eqno (2.19)$$
$$=\int \cD A_\mu^\perp \cD \omega \det (-\Delta )^{\frac 12}_S
\exp (i<A^\perp +\partial \omega , A^\perp +\partial \omega>)=$$
$$=\int \cD A_\mu \exp (i<A,A>)=1$$
where we used the definition (2.16) of the covariant measure for
scalar field and the property $<A^\perp ,\partial \omega >$$=0$.

The gauge transformations $\omega$ can be parametrized by the
gauge fixing functions $ \chi$ , $\tilde \chi $. The Jacobian factor
appearing due to the change of variables $\omega \to$$(\chi ,
\tilde \chi )$ is $[\tilde J \det \{ \chi ,\Phi \} ]^{-1}$.
In the adapted coordinates $( A^\perp ,\chi ,\tilde \chi )$
the path integral (2.17) takes the form
$$Z=\int \cD A^\perp \cD \chi \cD \tilde \chi
\det (-\Delta )_S^{\frac 12} \exp [-iS(A(A^\perp , \chi ,
\tilde \chi ) )] \delta (\chi ) \delta (\tilde \chi )
\eqno (2.20)$$
The action for electrodynamics can be rewritten as
$$S(A)=\frac 12 \iMi [A_\mu (g^{\mu \nu}\Delta
-\nabla^\mu \nabla^\nu +R^{\mu \nu})A_\nu ] \eqno (2.21)$$
It is interesting to note that the operator in the action
(2.21) maps four-vectors to transversal four-vectors:
$$\nabla_\mu (g^{\mu \nu}\Delta
-\nabla^\mu \nabla^\nu +R^{\mu \nu})A_\nu =0 \eqno (2.22)$$
The path integral (2.20)is easily evaluated giving
$$Z=\det (-g^{\mu \nu} \Delta -R^{\mu \nu})_\perp^{-\frac 12}
\det (-\Delta )_S^{\frac 12} .\eqno (2.23)$$
The subscript $\perp$ means that the determinant is evaluated
on the space of transversal four-vectors. This is the well known
covariant expression for the path integral of QED. We derived
it starting from the canonical Hamiltonian expression. Hence
it is both unitary and covariant.

Consider the expresssion for the path integral in terms of
physical degrees of freedom [13]
$$Z_{\rm PDF} =\det (-g^{\mu \nu}\Delta -R^{\mu \nu})_T^{-\frac 12}
\eqno (2.24)$$
where the determinant is restricted to transversal three-vectors. This
expression can be obtained from (2.10) by formal integration over
$\lambda$, $A_i$ and $P^k$. However, one should suppose that all
the measures are Gaussian and integration do not give rise to any
new factors. The expression (2.24) can not be obtained if one uses
only the covariant integration measures for four-vectors and
scalars. The non-covariance of the path integral (2.24) is
manifested through half-integer powers of the proper time in the heat
kernel expansion (1.1) on de Sitter space [11]. Other effects of non-
covariance of (2.24) will be described in the next section.
Note here that (2.24) do not follow from canonical quantization of
electrodynamics if one does not make any additional assumptions.

Let us extend the integration in (2.23) to all vector fields.
This is done with the help of equation
$$\det (-g^{\mu \nu}\Delta -R^{\mu \nu} )_{\parallel}=
\det (-\Delta )_S \eqno (2.25)$$
Using orthogonality of the decomposition (2.18) and the fact the
the operator under determinant do not mix longitudinal and
transversal four-vectors we can represent the path integral (2.23)
in the following form
$$Z=\det (-g^{\mu \nu} \Delta -R^{\mu \nu})_V^{-\frac 12}
\det (-\Delta )_S \eqno (2.26)$$
The subscript $V$ means all four-vector fields.

Consider now an important particular case of static space-time,
$M^4=M^3\times T^1$. For the sake of convenience we shall work
with the Euclidean signature space-time. On a static background
the configuration space of all vector fields can be decomposed
in a direct sum
$$V=V_T \oplus V_L \oplus V_0 \eqno (2.27)$$
where $V_T$ and $V_L$ are spaces of transversal and longitudinal
vector fields respectively, $V_0$ is the space of vectors having
only zeroth component. Let the eigenvalues of the scalar
three-Laplacian be $-\lambda_l$ while $l=0$ corresponds to
constant field , $\lambda_0=0$ (if constant mode is allowed by
boundary conditions on $M^3$). The eigenvalues of the scalar
four-Laplacian $\Delta_S$ are $-k^2r^{-2}-\lambda_l$, $k=$
$0,\pm 1, \pm 2 ,...$, $r$ is the radius of $T^1$. After some
algebra one can obtain that the eigenvalues of the operator
$g^{\mu \nu}\Delta +R^{\mu \nu}$ on $V_L$ and $V_0$ are also
$-k^2r^{-2}-\lambda_l$, but on $V_L$ the $l=0$ harmonics should
be excluded since they do not generate longitudinal fields.
Hence,
$$\det (-g^{\mu \nu}\Delta -R^{\mu \nu} )_L=
\prod_k \prod_{l \ne 0} (r^{-2}k^2+\lambda_l)$$
$$\det (-g^{\mu \nu}\Delta -R^{\mu \nu} )_0=
\prod_{(k,l) \ne 0} (r^{-2}k^2+\lambda_l)$$
$$\det (-\Delta  )_S=
\prod_{(k,l) \ne 0} (r^{-2}k^2+\lambda_l) \eqno (2.28)$$
In the last two equations $l$ and $k$ can not be zero
simultaneously. Substituting (2.28) in (2.26) we obtain that
contribution of scalar ghost is cancelled by contribution
of $V_L$ and $V_0$ fields up to the factor
$$\prod_{k \ne 0} (r^{-2}k^2)^{\frac 12} \eqno (2.29)$$
It is easy to see that the expression (2.29) is just the
Jacobian factor $\tilde J$ appearing due to the condition
(2.11).
$$Z=\tilde J^{\frac 12} det (-g^{\mu \nu}\Delta-R^{\mu \nu}
)_T^{-\frac 12}=\tilde J^{\frac 12} Z_{\rm PDF} \eqno (2.30)$$
We see that on static background the non-covatiant expression
(2.24) should be modified only by the factor $\tilde J^{\frac 12}$.
This factor corresponds to one-dimensional field theory and
describes less than one degree of freedom in four dimensions.
However, this factor contributes to the energy-momentum tensor.
Note that even on non-static background the inclusion
of $\tilde J$ leads to cancellation of manifestly
non-covariant terms in the heat kernel expansions (1.1)
and (1.2) [20,23].
If $M^3=R^3$ the scalar modes with $l=0$ are excluded by
 conditions at spatial infinity and the expression is
 correct without any modifications.

\newpage
{\bf 3.Examples}
\newline
3.1. QED on four-torus

Consider electrodynamics on $T^4$. the Euclidean effective
action corresponding to the path integral (2.24) over physical
degrees of freedom is
$$-\log Z_{\rm PDF}=\frac 12 [2\sum_{n_0}
{\sum_{n_1,n_2,n_3}}'
\log (
\frac {n_1^2}{r_1^2}+
\frac {n_2^2}{r_2^2}+
\frac {n_3^2}{r_3^2}+
\frac {n_0^2}{r_0^2} ) +3{\sum_{n_0}}' \log \frac {n_0^2}{r_0^2}
] \eqno (3.1)$$
where prime means that all summation indices can not be
equal to zero simultaneously. The $r_0,...,r_3$ are radii of
$T^4$. The coefficient 2 before the first sum reflects presence
of two helicity components for any non-zero vector $(n_1,
n_2,n_3)$. The coefficient 3 before the second sum is due to
three independent components of a transversal three-vector for
$n_1=$$n_2=$$n_3=0$. The expression (3.1) obviusly depends
on choice of the $x_0$ direction among coordinates on $T^4$.

The Jacobian factor $\tilde J$ (2.29) has the form
$$\log \tilde J^{\frac 12}=\frac 12 {\sum_{n_0}}'
\log \frac {n_0^2}{r_0^2} \eqno (3.2)$$
Now we can write the covariant effective action corresponding to
the path integral (2.30)
$$-\log Z=
{\sum_{n_0,n_1,n_2,n_3}}'
\log (
\frac {n_1^2}{r_1^2}+
\frac {n_2^2}{r_2^2}+
\frac {n_3^2}{r_3^2}+
\frac {n_0^2}{r_0^2} ) \eqno (3.3)$$
The same expression can be obtaine directly from (2.26). The
effective action (3.3) is symmetric under re-labeling coordinates
on $T^4$ as it should be.
\vspace {3mm}
\newline
3.2.Vector field on four-sphere

Consider a massless field theory on sphere $S^{d+1}$ with
amximally symmetric metric. If we apply a symmetry-preserving
analitical regularization, e.g. the zeta-regularization, the
regularized vacuum energy momentum tensor $T_{\mu \nu}^{reg}$
is proportional to $g_{\mu \nu }$ [4] because the $g_{\mu \nu}$
is the only symmetric invariant tensor on $S^{d+1}$.
$$T_{\mu \nu}^{reg}=\frac 1{d+1} g_{\mu \nu} (T^{reg})^\rho_\rho
\eqno (3.4)$$
The right hand side of (3.4) is proportional to the conformal
anomaly, which is finite in zeta-rugalarization. Thus the
left hand side should be finite too.
For the simplest case of scalar theory on $S^2$ the relation
(3.4) is established in the Appendix A.

In this subsection we shall demonstarte that the energy-
momentum tensor for $Z_{\rm PDF}$ is divergent on $S^4$. This means
that the rotational symmetry is broken.

The metric on $S^4$ has the form
$$ds^2=dx_0^2+\alpha \sin^2 x_0 d\Omega^2 \eqno (3.5)$$
where $d\Omega^2$ is the metric on unit $S^3$. We
introduced a real squashing parameter $\alpha$. The
$\alpha =1$ corresponds to unit roound $S^4$.

By definition
$$T_{\mu \nu} =\frac 2{\sqrt g}
\frac {\delta W}{\delta g^{\mu \nu}}, \quad
W=-\log Z \eqno (3.5)$$
Hence
$$\frac {\delta W}{\delta \alpha} \vert_{\alpha =1}
=-\frac 12 \int d^4 x T_{ij}g^{ij} \sqrt g
=-\frac 12 <T_i^i> \eqno (3.6)$$
The $T^i_i$ in the right hand side of (3.6) referes to the
round unit $S^4$. According to (3.4) it should be finite
if the $O(4)$ symmetry is unbroken.

Let us decompose the "physical" fields $A^T$ in series of
transversal vector harmonics $Y_{(l)}$ on $S^3$,
$l=1,2,...$ (see e.g. [31])
$$A_i^T=\sum_{(l)} f_{(l)}(x_0) Y_{(l)i}(x_k) \eqno (3.7)$$
Substitution of (3.7) in the eigenvalue equation
$(g^{ik}\Delta+$$R^{ik})A^T_k=$$\lambda A^{Ti}$ gives an
ordinary differential equation for $f_{(l)}$
$$[\partial_0^2 +ctg(x_0)\partial_0-
\frac {(l+1)^2}{\alpha \sin^2 (x_0)} ]f_{(l)}=
\lambda f_{(l)} \eqno (3.8)$$
To solve this equation we use the method of the paper [29].
After the change of variables
$$f=h \sin^b(x_0), \quad b=(l+1)\alpha^{-\frac 12}$$
$$z=\frac 12 (\cos (x_0) +1) \eqno (3.9)$$
the equation (3.8) takes the form
$$z(z-1)h''+(1+c)(z-\frac 12 )h'+eh=0,$$
$$e=b(b+1) +\lambda , \quad c=2b+1 \eqno (3.10)$$
Prime denotes differentiation with the respect to $z$.
According to the general prescription let us express
$h$ as power series [32]
$$h(z)=\sum_{k=0} a_kz^k \eqno (3.11)$$
By substituting (3.11) in (3.10) we get the recurrent
condition for the coefficients $a_k$
$$a_{k+1}=
\frac {k(k-1)+(1+c)k+e}{(k+1)(k+\frac 12 (c+1))}
\eqno (3.12)$$
The denominator of (3.12) is positive for all $k$.
The polynomial eigenfunctions $h_k$ can be found by imposing
the condition on the numerator to be equal to zero for some $k$.
We obtain the eigenvalues
$$\lambda_{(l)k}=-k^2-k(2 \frac {(l+1)}{\sqrt \alpha} +1)
-\frac {(l+1)}{\sqrt \alpha} (\frac {(l+1)}{\sqrt \alpha} +1)
\eqno (3.13)$$
The degeneracies of $\lambda_{(l)k}$ are the same as of
corresponding three-dimensional transversal vector spherical
harmonics
$$D_l^T=2l(l+2) \eqno (3.14)$$
In the limit $\alpha =1$ we obtain correct eigenvalues of the vector
Laplacian on $S^4$, $\lambda=-((k+l)+1)(k+l+2)=$$-p(p+1)$,
$p=k+l+1=2,3,$...The functions $h_k$ become associated Legandre
polynomials.

The effective action corresponding to the path integral
$Z_{\rm PDF}$ (2.24) is
$$W_{\rm PDF}=\frac 12 \sum_{l,k} D_l^T \log
\lambda_{(l)k}
\eqno (3.15)$$
In a framework of the zeta-regularization
$$-\frac 12 <T_i^i>=\frac {\delta W}{\delta \alpha}
\vert_{\alpha =1} =\frac 12 \lim_{s\to 0} \sum_{l,k}
D_l^T \left ( \frac {d \lambda_{(l)k}}{d\alpha}
\lambda_{(l)k}^{-(1+s)} \right )_{\alpha =1}=$$
$$=\frac 12 \lim_{s \to 0} \sum_{p=2}^{\infty}
\sum_{l=1}^{p-1}
\frac {(p+\frac 12 )(l+1) D_l}{[p(p+1)]^{1+s}}=$$
$$=\frac 14 \lim_{s \to 0} \sum_{p=2}^{\infty}
\frac {(p^2(p+1)^2 -2p(p+1)-24)(p+\frac 12 )}{
[p(p+1)]^{1+s}} \eqno (3.16)$$
In its'more rigorous version the zeta-regularization should be
applied directly to the effective action. The result for
$<T_i^i>$ is the same up to the multiplier $\Gamma (s)s$
which is unity in the limit $s\to 0$.

The last line of the equation (3.16) can be interpreted as
a value at the point $y=1$ of the generalized zeta-function $\zeta (y)$
of the operator with eigenvalues $p(p+1)$ and the degeneracies
given by the numerator. We can use the fact that the residue of
$\Gamma (y) \zeta (y)$ at $y=1$ is equal to the coefficient
before $t^{-1}$ in the heat kernel-like expansion
of the following sum for small $t$
$$\sum_{p=2}^{\infty} \frac 14 (p^2 (p+1)^2-2p(p+1)-24)
(p+\frac 12 ) \exp (-t p(p+1)) \eqno (3.17)$$
The asymptotic behavior of (3.17) can be easily find using the
equation [10]
$$\sum (p+\frac 12 ) \exp (-tp(p+1))=\frac 1{2t} +O(t^0)
\eqno (3.18)$$
This gives
$$\Gamma (1) {\rm Res}_{y=1} \zeta (y)=-3\ne 0 \eqno (3.19)$$
Hence the energy-momentum tensor
$$-\frac 12 <T^i_i>=\lim_{s \to 0} \zeta (1+s) \eqno (3.20)$$
is divergent in zeta-regularization. This means that the $O(4)$
symmetry is broken in the theory described by the path integral
$Z_{\rm PDF}$. One can also verify that the Jacobian factor
$\tilde J$ do not contribute to $<T_i^i>$. Thus the difficulties
of the approach based on the integration over physical
degrees of freedom can not be resolved by simply multiplying
by $\tilde J$ as it was in the static case.

\vspace {5mm}
3.3 Normalizability of the wave function of the Universe

Recently [33] a criterion of the normalizability of the
wave function of the Universe was suggested. It reads
$$C>-1 \eqno (3.21)$$
where $C$ is the sum of anomalous scalings of all
the fields on the Euclidean four-sphere $S^4$.
In the paper [33] the contribution of vector fields was
computed in a framework of an approach based on
integration over physical degrees of freedom, which
corresponds to the path integral $Z_{\rm PDF}$ (see Ref.[11])
$$C_1^{\rm PDF}=-\frac {16}{45} \eqno (3.32)$$
{}From the above discussion it is clear that the anomalous
scaling (3.22) should be replaced by that coming from the
covariant path integral (2.26) [11]
$$C_1=\frac {59}{45} \eqno (3.23)$$
Note that another value of $C_1$ in covariant approach exists
in the literature [9,10]. The difference is due to miscounting
of ghost zero modes [11].

On the same grounds we prefer to use covariant path integral
for gravitational field. The corrrespondin scaling behavior
is [7]
$$C_2=\frac {329}{45} \eqno (3.24)$$
The computations in terms of physical degrees of freedom [8]
give $C_2^{\rm PDF}=-\frac {661}{45}$. For a review of
different approaches to quantum gravity on $S^4$ see [19].
For non-gauge scalar and Dirac spinor fields the values
of scaling behavior are unambiguous and well known (see e.g.
[4,10])
$$C_0=\frac {29}{90} -4\xi +12\xi^2,\quad C_{\frac 12}=
\frac {11}{90} \eqno (3.35)$$
where $\xi$ is the standard parameter of scalar coupling
to gravity.

For the system consisting of graviton, three generations of
fermions, gauge fields of the standard model and one complex
Higgs doublet we obtain
$$C=\frac {325}{12} +48 \xi^2 -16 \xi \eqno (3.26)$$
instead of the non-covariant result [34]
$$C^{\rm PDF}=-\frac {179}{12} +48 \xi^2 -16 \xi \eqno (3.27)$$
We see that the criterion (3.21) is fulfilled by (3.26) for
both minimal $(\xi =0)$ and conformal ($\xi =\frac 16$) coupling,
while the $C^{\rm PDF}$ do not satisfy this criterion.
The covariant path integral gives more optimistic predictions.

\newpage
{\bf 4. QED in a bounded region}

One of the most interesting applications of quantum field theory
on manifolds with boundaries is related to quantum cosmology
and the Hartle-Hawking wave function of the Universe.
In the semiclassical approximation the contribution of the
electromagnetic field to the wave function of the Universe
is given by the Euclidean path integral [35]
$$Z=\int \cD \mu (A) \exp (-S(A)) \eqno (4.1)$$
where the fields $A$ satisfy some conditions on the boundary
$\partial M$ of gravitational instanton $M$. The measure
$\cD \mu (A)$ includes all Faddeev-Popov determinants.

As in the case of unbounded manifolds the computations in terms
of physical degrees of freedom [13] and in terms of four-vector
fields [14,15] give different results [5,18].
A part of this contradictions
was due to a mistake in analytical formula for heat kernel
expansion in the case of mixed boundary conditions (see [23,24]).
However, even if correct formula is used the two above mentioned
approaches still disagree [24]. From our point of view this
difference originates from the fact that the Gaussian measure
on the space of physical variables is really non-covariant and
should be replaced by another measure obtained by reduction
of standard measure on the space of four-vector fields to
the subspace of physical variables. An interesting explanation
why these measures are different was suggested recently [21,22].
We postpone for a while discussion of this suggestion. In this
section we shall derive the covariant Moss and Poletti path
integral [14,15,27] starting from the canonical formalism. This will
prove unitarity of the path integral [14]. Next we shall
derive the same path integral from the manifestly covariant
expression by means of the so-called geometric approach to
quantization of gauge theories [36].
\vspace {5mm}
\newline
4.1 Canonical path integral

Let the manifold $M$ admit a coordinate system such that
$$ds^2=dx_0^2+g_{ik} dx^i dx^k \eqno (4.2)$$
and the boundary $\partial M$ corresponds to a constant
value of the Euclidean time $x_0$. All canonical conjugate
pairs will be defined with respect to $x_0$ variable.

To make physical consideration more transparent we consider
a simplified case
$$g_{ik}(x_0,x_j)=c^2(x_0) \tilde g_{ik}(x_j) \eqno (4.3)$$
We suppose also that the metrics $g_{\mu \nu}$ and
$\tilde g_{ik}$ satisfy the Einstein equations. This ensures
that the Ricci tensors are covariantly constant. All these
restrictions, however, do not exclude any physically
interesting example, such as disk or a spherical segment.
We shall use notation $\Gamma (x_0)=$$c^{-1}\partial_0 c$.

Let us start with the path integral [35,13] over physical
components
$$Z=\int (\cD A_i^T)_D \exp (-S(A^T)) \eqno (4.4)$$
where integration is performed over transversal three-vectors
satisfying the Dirichlet boundary condition
$$A^T_i \oB \eqno (4.5)$$
Unlike to Ref .[13] we will not suppose that the
measure $(\cD A^T_i)_D$ is Gaussian. Instead we enlarge
the integration domain selfconsistently in order to
obtain integrals over four-vectors. We shall omitt
the steps which are identical to that of sec. 2. Special
attention will be paid to boundary conditions.

According to the general method [1,2] the path integral (4.4)
should be considered as a result of integration over momenta
in the reduced phase space path integral. Let us recall the
first order form of the Euclidean action for QED
$$S=\int d^4x \sqrt g [P^i\partial_0A_i+
\frac 12 (\frac 12 F_{ik}F^{ik}-P_iP^i)+
A_0\ ^(3)\nabla_i P^i ] \eqno (4.6)$$
This action is written in terms of independent variables
$P^i=F^{0i}$ and $A_i$, $A_0$ is the Lagrange multiplier.
The integration by parts over $x_i$ is always allowed
because either spatial slices are
are closed manifolds
or all the fields
decay rapidly at spatial infinity.

Boundary conditions for the reduced phase space momenta
$P^{Tj}$ are not arbitrary. They are specified by the
boundary conditions (4.5) for $A_i^T$ through the equation
$$P^i=g^{ik}\partial_0 A_k \eqno (4.7)$$
Note that the eq. (4.7) does not mean that $A_i$ and
$P^k$ become dependent variables. This equation rather
states a map between two functional spaces which enables
us to interpret $A_i$ and $P^k$ as cojugate variables.
As it is demonstrated in the Appendix B, the relation (4.7)
leads to the following boundary condition on $P^{Ti}$
$$(\partial_0 +3\Gamma )P^{Ti} \oB  \eqno (4.8)$$
In order to be able to use covariant definition of
the path integral measure we are to extend the phase
space in such a manner to obtain integration over
unconstrained four-vector fields. Boundary conditions
for the non-physical components and ghosts should
be choosen self-consistently.

The most natural choice for spatial components
$A_i$ and $P^i$is to extend the Dirichlet and
Neumann boundary conditions (4.5) and (4.8) to
all three-vector fields.
$$A_i \oB ,\quad
(\partial_0 +3\Gamma )P^{i} \oB \eqno (4.9)$$
To generate the constraint $\Phi =-^{(3)}\nabla_iP^i$$=0$
the Lagrange multiplier should belong to the same
functional space and hence satisfy the same boundary
condition as $\Phi$, namely
$$(\partial_0 +3\Gamma )A_0 \oB \eqno (4.10)$$
The boundary conditions for the ghosts are the same as
for the parameter $\omega$ of gauge transformation
which map the space defined by (4.9) and (4.10) in
itself. One can demonstrate that these conditions are
(see Appendix B)
$$\omega \oB \eqno (4.11)$$

Note that the equations (4.9)-(4.11) describe a particular case
of more general BRST-invariant boundary conditions
proposed by Moss and Poletti [14]
$$A_i \oB , \quad (\nabla_0 +k_i^i)A_0 \oB ,\quad
\omega \oB \eqno (4.12)$$
where $k_{ij}$ is the second fundamental form of the
boundary.

The last ingredient we need for the case of compact spatial
sections is the additional gauge fixing condition (2.11).
It is obvious, that to obtain covariant configuration
space one should extend boundary conditions (4.12) to
spatially constant fields $A_0$ and $\omega$. The operator
$1\otimes \{ \Phi , \chi \} +$$(-\nabla_0^2) \otimes 1$,
where the first multiplier acts on $x_i$-independent fields
and the second one only on the fields with non-trivial
$x_i$-dependence, is non-degenerate ghost operator.

Boundary conditions for the fields $P^0$ and $a$ are not
significant because these fields are immediately integrated
out. The covariance of the obtained path integral will
be stated independently in the next sub-section.

Now we are able to write the Euclidean path integral in
the form (see (2.17))
$$Z=\int \cD A_\mu \delta (\chi ) \delta (\tilde \chi )
\tilde J \exp (-S(A_\mu)) \eqno (4.13)$$
where the fields $A_\mu$ and ghosts in the Jacobian factors
satisfy the above defined boundary conditions.
To obtain Lorentz gauge path integral we shall use the
Faddeev-Popov trick. Let us insert the unity in the
path integral (4.13)
$$1=\int \cD \omega \det (-\Delta ) \delta
(\nabla^\mu A_\mu (\omega )), \quad A_\mu (\omega )=
A_\mu +\partial_\mu \omega \eqno (4.14)$$
with $\omega$ satisfying Dirichlet boundary condition.
After change of variables $A_\mu \to$$A_\mu (-\omega )$
and integration over $\omega$ with the help of another
representation of unity
$$1=\int \cD \omega \det \{ \Phi , \chi \}
\tilde J \delta (\chi (A(-\omega )))
\delta (\tilde \chi (A(-\omega ))) \eqno (4.14a)$$
we arrive at
$$Z=\int \cD A_\mu \det (-\Delta ) \delta (\nabla^\mu A_\mu )
\exp (-S(A_\mu )) $$
$$=\int \cD A_\mu^\perp \det (-\Delta )^{\frac 12 }
\exp (-S(A_\mu^\perp )) \eqno (4.15)$$
for derivation of (4.14) and (4.14a) it is essential that
Lorentz gauge condition fixes the gauge freedom completely,
as well as the conditions $\chi$ and $\tilde \chi$. To
obtain the last line of equation (4.15) we integrated over
longitudinal fields using the fact that due to conditions
(4.12) transversal fields are orthogonal to longitudinal
ones with the respect to ordinary scalar product without
surface terms. Of course, all the fields in (4.15) again
satisfy conditions (4.12). The expression (4.15) is the
Lorentz gauge path integral proposed by Moss and Poletti.
Here we derived it directly from the canonical
reduced phase space quantization. This supports conclusion
made previously on the basis of BRST-invariance [14] that
the path integral (4.15) describes unitary theory.

\vspace {5mm}

4.2. Covariance of the path integral

The boundary conditions (4.12) have an advantage [23,25]
that the Hodge-de Rham decomposition (2.18) is orthogonal
with the respect to ordinary scalar product in the space
of vector fields without surface terms and the Laplace
operator is self-adjoint. However, these conditions are
not manifestly covariant. Different components of $A_\mu$
obey different types of boundary conditions. In this
subsection we re-obtain path integral (4.15) , (4.12)
by means of manifestly covariant procedure in a framework
of the so-called geometric approach to quantization of
gauge theories [36,30,37,38].

The starting point is the path integral without gauge
fixing terms
$$Z=\frac 1{{\rm vol}\ G} \int \cD A_\mu
\exp (-S(A_\mu )) \eqno (4.16)$$
where vol $G$ is infinite volume of the gauge group.
We assume that all the components of $A_\mu$ satisfy
Dirichlet boundary condition
$$A_\mu \oB \eqno (4.17)$$
and the gauge group is spanned by gradient transformations
with the parameter $\omega$ satisfying Dirichlet boundary
too
$$\omega \oB \eqno (4.18)$$
The boundary conditions (4.17) and (4.18) are manifestly
covariant.

The action $S(A_\mu )$ is independent of pure gauge
degrees of freedom. We can integrate over longitudinal
fields in (4.16) thus cancelling the volume of the
gauge group. To this end let us use the Hodge-de Rham
decomposition
$$A_\mu =A_\mu^\perp +\partial_\mu \omega \eqno (4.19)$$
$$\nabla^\mu A_\mu^\perp =0 \eqno (4.20)$$
Due to the condition (4.18) this decomposition
is orthogonal with the respect to ordinary inner
product without surface terms.

{}From (4.17), (4.18) and (4.19) we see that spatial
components of $A^\perp$ satisfy the Dirichlet
condition
$$A_i^\perp \oB \eqno (4.21)$$
As it was explained in the previous subsection
(see also Appendix B), the condition (4.21)
together with (4.18) and (4.19) lead to
$$(\partial_0 +k_i^i ) A_0^\perp \oB \eqno (4.22)$$
To verify consistency of the procedure, let us
substitute (4.21) and (4.22) in the gauge
condition on the boundary
$$\nabla^\mu A_\mu^\perp \vert_{\partial M} =
(\partial_0 +k_i^i )A_0^\perp \vert_{\partial M}
+\ ^{(3)}\nabla^i A_i^\perp \vert_{\partial M}=0
\eqno (4.23)$$
Due to the orthogonality property the path integral
measure can be represented in the form
$$\cD A_\mu =\cD A_\mu^\perp \cD \omega
\det (-\Delta )^{\frac 12} \eqno (4.24)$$
Substituting (4.24) in (4.16) and integrating over $\omega$
with the help of identity vol$G$$=\int \cD \omega$ we arrive
at
$$Z=\int \cD A_\mu^\perp \det (-\Delta )^{\frac 12}
\exp (-S(A^\perp)) \eqno (4.25)$$
with the boundary conditions (4.18) , (4.21) and (4.22).
This is exactly the Moss and Poletti path integral (4.15).
We have demonstrated that their result is covariant.
\newpage
{\bf 5. Conclusions}

Let us summarize in short the main results of this paper.
We analized the reduced phase space quantization procedure
on curved background. It was demonstrated that the Hamiltonian
quantization plus manifestly invariant measure lead to path
integral identical to that obtained in the gauge fixed
approach. This procedure was also applied to manifolds with
boundaries. It was shown that the Faddeev-Popov trick lead
to the Moss and Poletti [14] boundary conditions. The same
boundary conditions were also obtained by using manifestly
covariant method based on geometric approach to quantization
of gauge theories. With the example of QED on four-sphere
we demonstrated that the use of Gaussian measure on the
space of physical degrees of freedom leads to breakdown
of the $O(4)$ rotational symmetry. Some important particular
cases were also analised. A criterion of normalizability of
the wave function of the Universe was applied to the covariant
path integral with field content identical to that of the
Standard Model.

The main problem we addressed to in this paper was as follows.
There are two formulations of path integral for gauge theories
on curved background and on manifolds with curved boundaries.
The one formulation is manifestly covariant, the other is
manifestly unitary. These two formulations disagree. The roots
of this disagreement are in use of different path integral
measures. If one uses covariant measure in both cases, one
gets identical results. The covariant path integral appeared
to be also unitary because it was obtained from the Faddeev
reduced phase space integral.

The other question is why the two measures are different.
The results of this paper support the point of view [21,22]
that the reason is that the 3+1 decomposition is ill-defined on
curved background and on some manifolds with boundaries.
However it is yet unclear whether this effect is pure
topological and is entirely due to the presence of singular
point-like spatial slices. Anyhow, this problem deserves
further investigation. One should also check up independently
unitarity of the covariant path integral and find out which
fundamental symmetries are broken in the approach based on
physical degrees of freedom on manifolds with boundaries.

The results of this paper can be also interesting from the
point of view of equivalence of different quantization techniques
fro gauge theories (see e.g. [30,37,39]). It would be important to
derive covariant path integral measure from the action principle
and canonical quantization.

\vspace {5mm}

{\bf Acknowledgements}

The author is grateful to Giampiero Esposito for discussions
and comments. He would also like to thank Professor Abdus
Salam, IAEA and UNESCO for hospitality at the International
Centre for Theoretical Physics, Trieste. This work was supported
by the Russian Foundation for Fundamental Studies, grant
93-02-14378.
\newpage
{\bf Appendix A}

As a simple exmple of application of the technique of sec. 3.2
to covariant case consider minimaly coupled scalar field on
the two-sphere $S^2$. Introduce a deformed metric
$$ds^2=dx_0^2 +\alpha \sin^2 (x_0) dx_1^2 \eqno (A.1)$$
The scalar Laplace operator reads
$$\Delta = (\partial_0+ {\rm ctg}(x_0))\partial_0
+\frac 1\alpha \partial_1^2 \eqno (A>2)$$
The spectrum of $\Delta$ can be found by the same
manipulations as in the sec. 3.2.
$$\lambda_{k,l}=-k^2- k( 2
\frac {\abL}{\sqrt \alpha} +1)-\frac {\abL}{\sqrt \alpha}
(\frac {\abL}{\sqrt \alpha} +1)$$
$$k=0,1,..., \quad l=0,\pm 1,\pm 2,... \eqno (A.3)$$
The $\abL$ is the absolute value of $l$. We have for the
energy-mamentum tensor
$$-\frac 12 <T^1_1>=\frac {\delta W}{\delta \alpha}
\vert_{\alpha =1}=-\frac 12 \lim_{s\to 0}
\sum_{k,l} \frac {k \abL +l^2+\frac 12 \abL}{[(k+\abL )
(k+\abL +1)]^{1+3}}=$$
$$=-\frac 12 \lim_{s\to 0} \sum_{n=1}^\infty
\sum_{l=1}^n \frac {(2n+1)l}{[n(n+1)]^{1+s}} =
-\frac 14 \lim_{s\to 0} \sum_{n=1}^\infty
\frac {(2n+1)(n+1)n}{[n(n+1)]^{1+s}}=$$
$$=-\frac 14 \lim_{s\to 0} \sum_{n=1}^\infty
\frac {2n+1}{[n(n+1)]^s} =-\frac 14 \zeta (0)
\eqno (2.4)$$
where $n=\abL +k$. The $\zeta (0)$ is just the zeta-function
of the operator (A.2) for $\alpha =1$. The crucial difference
between (A.4) and (3.16) is that now the multiplier $n(n+1)$
in the numerator cancells the one in denominator. Thus the
corresponding term in the heat kernel expansion before zeroth
power of proper time is equivalent to the residue of
$\Gamma (y) \zeta (y)$ at the point $y=0$ (and not at $y=1$ as in
(3.16)). At $y=0$ the $\Gamma$-function itself has a pole.
Hence the $\zeta (y)$ remains finite at this point. The value
of residue is well known (see e.g. [10]); it gives $\zeta (0)=$
$\frac 13$. It is instructive to compare (A.4) with
expression for the conformal anomaly or one-loop scaling
behaviour.
$$<T_\mu^\mu >=\lim_{s\to 0} \sum_{k,l}
\frac {-\lambda_{k,l}}{[-\lambda_{k,l}]^{1+s}} =
\lim_{s\to 0} \sum_{n=1}^\infty \sum_{l=-n}^n
\frac 1{[n(n+1)]^s} =$$
$$=\lim_{s\to 0}  \sum_{n=1}^\infty
\frac {2n+1}{[n(n+1)]^s} =\zeta (0) \eqno (A.5)$$
We have demonstrated that in zeta-regularization
$$<T_\mu^\mu >=2<T_1^1>=\zeta (0) \eqno (A.6)$$
and the energy-momentum tensor is finite as it should
be for a covariant theory on $S^2$ with unbroken
$O(2)$-symmetry.
\vspace {5mm}
\newline
{\bf Appendix B}

This Appendix contains necessary information on eigenmodes
of the Laplace operator on manifolds with boundaries admitting
metric described by eqs. (4.2) and (4.3). The Laplace
operator acting on scalar and vector fields has the form
$$\Delta \phi =(\partial_0^2+3\Gamma \partial_0
+\ ^{(3)}\Delta )\phi ,$$
$$(\Delta A)_0=[\partial_0^2+3\Gamma \partial_0
+\ ^{(3)}\Delta -3\Gamma^2 ]A_0 -2 \Gamma
\ ^{(3)}\nabla^i A_i , \eqno (B.1)$$
$$(\Delta A)_i=[\partial_0^2+\Gamma \partial_0
+\ ^{(3)}\Delta -3\Gamma^2 +\Gamma (\partial_0 \Gamma )]A_i
+2\Gamma \ ^{(3)}\nabla_i A_0$$
Let us remind that
$$g_{ik}=c^2(x_0) \tilde g_{ik} (x_j); \quad
\Gamma =\frac {\partial_0 c}c .\eqno (B.2)$$
Since we assumed that the $g$ and $\tilde g$ are Einsteinian,
the terms with Ricci tensors in vector operators simply lead
to constant shift of eigenvalues. We can ignore Ricci tensor
in what following.

The metric (4.2), (4.3), (B.2) admits separation of variables
in the operators (B.1). Let $\tilde \Delta$ be the
3-dimensional Laplace operator built up with metric
$\tilde g_{ik}$, $\ ^{(3)}\Delta =$$c^{-2}\tilde \Delta$.
Let us denote scalar and transverse vector eigenfunctions
of $\tilde \Delta$ as $Y_{(l)}^S$ and $Y_{(l)j}^T$. The
four-dimensional vector and scalar fields can be decomposed
in the following series of orthogonal harmonics
$$\phi (x_0,x_j)=\sum_{(l)} \phi_{(l)} (x_0) Y_{(l)}^S (x_j)
 , \quad
A_0 (x_0,x_j)=\sum_{(l)} a_{(l)} (x_0) Y_{(l)}^S (x_j) ,$$
$$A_i (x_0, x_j)=\sum_{(l)} v^T_{(l)}(x_0) Y^T_{(l)i}(x_j)
+\sum_{(k)} v^L_{(k)}(x_0)\partial_i Y^S_{(l) }(x_j)
\eqno (B.3)$$
Substituting (B.3) in the eigenvalue equations $\Delta_\phi
=\lambda \phi$ and $\Delta A_\mu =\lambda A_\mu$ we obtain
ordinary differential equations for $\phi_{(l)}$ and $v^T_{(l)}$
and pairs of coupled ordinary differential equations for
$a_{(k)}$ and $v_{(k)}$. It means that scalar and transversal
3-vector can be decomposed in sums of harmonics
which are eigenfunctions both $\Delta$ and $\tilde \Delta$.
Transversal 3-vectors decouple from the other fields.

As far as functional integral is concerned we can establish
boundary conditions for separate eigenmodes of the Laplace
operator [27].

First let us prove the equation (4.8)
$$(\partial_0 +3\Gamma )  P^{Ti}=
(\partial_0 +3\Gamma ) g^{ik} \partial_0 A_k^T=
g^{ik} (\partial_0 +\Gamma ) \partial_0 A_k^T=$$
$$=g^{ik} ( \Delta A_k^T - [\frac 1{c^2} \tilde \Delta
-3 \Gamma^2 +\Gamma (\partial_0 \Gamma) ]A_k^T \eqno (B.4)$$
where we used decoupling of $A^T$ from other vector harmonics
and the equation (B.1). Let $A_i^T$ be an eigenmode of $\Delta$
and $\tilde \Delta$. When the expression in the last line of
(B.4) vanishes term by term on the boundary if $A_i^T$
satisfies Dirichlet boundary condition (4.5).

Next we demonstrate that if the gauge parameter $\omega$ satisfy
the Dirichlet condition (4.11) when the pure gauge vector potential
$A_\mu =\partial_\mu \omega$ satisfy the conditions (4.9) and
(4.10). The Dirichlet condition (4.10) for $A_i$ is obvious.
Indeed, differentiation with respect to $x_i$ do not affect the
coefficient function $f_{(l)}(x_0)$which vanishes on the
boundary due to (4.11). For the zeroth component we have
$$(\partial_0 +3\Gamma )\partial_0 \omega =
\Delta \omega -\frac 1{c^2} \tilde \Delta \omega \eqno (B.5)$$
Since the harmonics of scalar field $\omega$ can be chosen
to be eigenfunctions on both $\Delta$ and $\tilde \Delta$, the
right hand side of (B.5) vanishes on the boundary term by term
due to the Dirichlet condition (4.11).

After obvious rearrangements in the above reasoning
on can also obtain the equation (4.22). Let us remind,
that on the boundary $k_{ij}=\Gamma g_{ij}$ if the
outward pointing normal vector is directed along positive
$x_0$ axis.
\newpage
{\bf References}
\newline
[1] L.D.Faddeev and V.N.Popov, Phys.Lett. 25B, 29 (1967).
\newline
[2] L.D.Faddeev, Teor. Mat. Fiz. 1, 3 (1969);

L.D.Faddeev and A.A.Slavnov, Gauge fields: Introduction to
quantum theory, Benjamin/Cummings, 1980.

L.D.Faddeev and R.Jackiw, Phys. Rev. Lett. 66, 1692 (1988).
\newline
[3] B.S.DeWitt, in Relativity, Groups and Topology, eds. C.DeWitt
and B.S.DeWitt, Gordon and Breach, NY, 1964.
\newline
[4] N.D.Birell and P.C.W.Davies, Quantum fields in curved space,
Cambridge Univ. Press, Cambridge, 1982;

I.L.Buchbinder, S.D.Odintsov and I.L.Shapiro, Effective action
in quantum gravity, IOP Publishing, Bristol, 1992.
\newline
[5] G.Esposito, Quantum gravity, quantum cosmology and Lorentzian
geometries, Springer, Berlin, 1992.
\newline
[6] G.W.Gibbons and M.J.Perry, Nucl. Phys. B146, 90 (1978);

S.M.Christensen and M.J.Duff, Nucl. Phys. B170, 480 (1980).
\newline
[7] T.R.Taylor and G.Veneziano, Nucl. Phys. B345, 210 (1990).
\newline
[8]] P.A.Griffin and D.A.Kosower, Phys. Lett. B233, 295 (1989).
\newline
[9] G.Shore, Ann. Phys. 117, 121 (1979).
\newline
[10] D.Birmingham, Phys. Rev. D36, 3037 (1987).
\newline
[11] D.V.Vassilevich, Nuovo Cimento 104A, 743 (1991).
\newline
[12] K.Schleich, Phys. Rev. D32, 1889 (1985).
\newline
[13] J.Louko, Phys. Rev. D38, 478 (1988).
\newline
[14] I.G.Moss and S.Poletti, Nucl. Phys. B341, 155 (1990).
\newline
[15] I.G.Moss and S.Poletti,Phys. Lett. B245, 355 (1990).
\newline
[16] A.O.Barvinsky, A.Yu.Kamenshchik and I.P.Karmazin, Ann. Phys. 219,
201 (1992); A.O.Barvinsky, A.Yu.Kamemshchik, I.P.Karmazin and
I.V.Mishakov, Class. Quantum Grav. 9, L27 (1992);

A.Yu.Kaminshchik and I.V.Mishakov, Int. J. Mod. Phys. A7, 3713 (1992).
\newline
[17] A.O.Barvinsky, Phys. Rep. 230, 237 (1993).
\newline
[18] G.Esposito, Nuovo Cimento 109B, 203 (1994); Class. Quantum.
Grav. 11, 905 (1994).
\newline
[19] D.V.Vassilevich, Int. J. Mod. Phys. A8, 1637 (1993).
\newline
[20] D.V.Vassilevich, Nuovo Cimento 105A, 649 (1992).
\newline
[21] A.Yu.Kaminshchik and I.V.Mishakov, Phys. Rev. D49, 816 (1994).
\newline
[22] G.Esposito, A.Yu.Kamenshchik, I.V.Mishakov and G.Pollifrone,
Euclidean Maxwell theory in the presence of boundaries, II,
DSF preprint 94/4, 1994, to appear in Class. Quantum Grav.;
Gravitons in one-loop quantum cosmology:
correspondence between covariant and non-covariant formalisms,
DSF preprint 94/14, 1994, to appear in Phys. Rev. D;

G.Esposito and A.Yu.Kamenshchik, Phys. Lett. B336, 324 (1994).
\newline
[23] D.V.Vassilevich, Vector fields on a disk with mixed
 boundary conditions, Preprint SPbU-IP-94-6, gr-qc/9404052.
 \newline
[24] I.G.Moss and S.Poletti, Phys.Lett. B333, 326 (1994).
\newline
[25] T.P.Branson and P.B.Gilkey, Commun. Part. Diff. Eqs. 15,
 245 (1990).
\newline
[26] D.J.Toms, Phys. Rev. D35, 3796 (1987);

E.Verdagner and P.van Nieuwenhuizen, Ann. Phys. 207, 77 (1991).
\newline
[27] H.Luckock, J.Math.Phys. 32, 1755 (1991).
\newline
[28] A.M.Polyakov, Phys. Lett. B103, 207 (1981).
\newline
[29] I.P.Grigentch and D.V.Vassilevich, Nuovo Cimento 107A,
 227 (1994).
\newline
[30] G.Kunstatter, Class. Quantum Grav. 9, 1469 (1992).
\newline
[31] M.A.Rubin and C.Ordonez, J. Math. Phys. 25, 2888 (1984);
26, 65 (1985).
\newline
[32] E.T.Wittaker and G.N.Watson, A Course of Modern Analysis,
Cambridge Univ. Press, Cambridge, 1927.
\newline
[33] A.O.Barvinsky and A.Yu.Kamenshchik, Class. Quantum Grav. 7,
L181 (1990).
\newline
[34] A.Yu.Kamenshchik, Phys. Lett. B316, 45 (1993).
\newline
[35] J.B.Hartle, Phys. Rev. D29, 2730 (1984).
\newline
[36] O.Babelon and C.-M.Viallet, Phys. Lett. B85, 246 (1979);
Commun. Math. Phys. 81, 515 (1981).
\newline
[37] Z.Bern, S.K.Blau and E.Mottola, Phys. Rev. D43, 1212 (1991);

P.O.Mazur and E.Mottola, Nucl. Phys. B341, 187 (1990).
\newline
[38] P.Ellicott, G.Kunstatter and D.J.Toms, Ann. Phys. 205, 70 (1991).
\newline
[39] P.Hajicek and K.Kuchar, Phys. Rev. D41, 1091 (1990);

C.R.Ordonez and J.M.Pons, Phys. Rev. D45, 3706 (1992);

R.J.Epp, G.Kunstatter and D.J.Toms, Phys. Rev. D47, 2724 (1993).

\end{document}